\documentstyle[psfig]{mn}

\begin{document}

\title[Spectral Evolution of SN 1998bw]{Spectral Evolution of the Peculiar Ic Supernova 1998bw}

\author[R. A. Stathakis et al.]{R. A. Stathakis,$^{1}$ B. J. Boyle,$^{1}$
 D. H. Jones,$^{2}$ M. S. Bessell,$^{2}$ T. J. Galama,$^{3}$ 
\newauthor
Lisa M. Germany,$^{2}$ M. Hartley,$^{1}$ D. M. James,$^{1}$ C. Kouveliotou,$^{4,5}$ I. J. Lewis,$^{1}$  
\newauthor
Q. A. Parker,$^{6}$ K. S. Russell,$^{1}$ E. M. Sadler,$^{7}$ C. G. Tinney,$^{1}$ J. van Paradijs$^{8,9}$ 
\newauthor
and P. M. Vreeswijk$^{8}$ \\ 
$^{1}$Anglo-Australian Observatory, P.O. Box 296, Epping, N.S.W. 1710,
 Australia \\
$^{2}$Research School of Astronomy and Astrophysics, The Australian National
University, Private Bag, Weston Creek, A.C.T. 2611, \\
Australia \\ 
$^{3}$Astronomy, MS 105-24, Robinson Lab, Caltech, Pasadena, CA 91125, USA \\
$^{4}$Universities Space Research Association \\
$^{5}$NASA Marshall Space Flight Center, ES-84, Huntsville, AL 35812, USA \\ 
$^{6}$Royal Observatory, Blackford Hill, Edinburgh EH9 3HJ, UK \\
$^{7}$School of Physics A29, University of Sydney, N.S.W. 2006, Australia \\
$^{8}$Astronomical Institute ``Anton Pannekoek", University of Amsterdam, \&
Center for High Energy Astrophysics, Kruislaan 403, \\
1098 SJ Amsterdam, The Netherlands \\
$^{9}$Physics Department, University of Alabama in Huntsville, Huntsville,
AL 35899, USA}

\maketitle
\begin{abstract}

SN 1998bw holds the record for the most energetic Type Ic explosion, one of the brightest radio supernovae and probably the first supernova associated with
a $\gamma$-ray burst. 
In this paper we present spectral observations of SN~1998bw observed in a cooperative monitoring campaign using the AAT, UKST and the SSO 2.3-m telescope. We investigate the evolution of the spectrum between 7 and 94 days after V-band maximum in comparison to  well-studied examples of Type Ic SNe in order to quantify the unusual properties of this supernova event. Though the early spectra differ greatly from observations of classical Ic SNe, we 
find that the evolution from the photospheric to nebular phases is slow but otherwise typical. The spectra differ predominantly in the extensive line blending and blanketing which has been attributed to the high velocity of the 
ejecta. We find that by day 19, the absorption line minima blueshifts are 10\% -- 50\% higher then other SNe and on day 94 emission lines are 45\% broader, as expected if the progenitor had a massive envelope. However, it is difficult to explain the extent of line blanketing entirely by line broadening, and we argue that additional contribution from other species is present, indicating unusual relative abundances or physical conditions in the envelope.

\vspace*{0.5cm}

\end{abstract}

\begin{keywords}
supernovae: general -- stars: evolution -- supernovae: general -- supernovae: individual: SN 1998bw -- gamma-rays: bursts 
\end{keywords}

\begin{table*}
\begin{minipage}{145mm}
\caption{Spectral observations of SN~1998bw.}
\begin{tabular}{ccccccccc}
UT Date & Age$^{1}$ & Telescope & Instrument & $\lambda$ Coverage & $\lambda$ Res. & Observer \\
 & (days) &  &  &  (\AA) & (\AA\ FWHM) & \\ 
 & & & & & & \\
98/05/19.7 & 7   & 2.3m & Nasmyth B Imager & 3600-10200 & 14 & Jones, Bessell \\

98/05/20.6 & 8   & UKST & FLAIR            & 4001-7151  & 13 & Russell, Parker \\

98/05/23.8 & 11  & 2.3m & Nasmyth B Imager & 3600-9990 & 14 & Jones, Bessell \\

98/05/23.8 & 11  & AAT  & 2dF              & 3600-7900  & 12 & Stathakis, Lewis \\

98/05/27.8 & 15  & AAT  & 2dF              & 4035-6263 & 5 & Lewis \\

98/05/29.7 & 17  & UKST & FLAIR            & 3950-7202 & 13 & Hartley, Parker \\

98/05/30.7 & 18  & UKST & FLAIR            & 3950-7202 & 13 & Hartley, Parker \\

98/05/31.8 & 19  & AAT  & RGO              & 3520-9300 & 5 & Stathakis, James \\
 
98/06/16.6 & 35  & UKST & FLAIR            & 5700-7560 & 12 & Hartley, Parker \\

98/06/26   & 45  & 2.3m & DBS              & 3850-7590 & 6 & Germany, Schmidt \\

98/08/14   & 94  & 2.3m & DBS              & 3800-7550 & 6 & Germany, Schmidt \\
 & & & & & & \\
\end{tabular}
\medskip
$^{1}$Age is given relative to the date of visual maximum, 1998 May 12.2 \cite{ga}. 
\end{minipage}
\end{table*}

\section{Introduction}

\subsection{SN/GRB association}

The suggestion that type Ic supernova (SN)~1998bw is the optical counterpart of $\gamma$-ray burst (GRB) 980425  has forced us to rethink the mechanics of both SNe and GRBs.
The association  of SN~1998bw and GRB980425 is supported by the extremely low
probability of a chance coincidence \cite{ga} and particularly by the peculiar observational characteristics of SN~1998bw. Patat \& Piemonte \shortcite{pa} have
classified SN~1998bw as a Type Ic supernova, since spectral lines due to helium, silicon and hydrogen are weak or absent in the early spectra. However, at $M_{B} = -18.88$ at maximum \cite{ga}, this object was three times brighter than the average SN Ic, and early spectra show extremely broad lines and unusual line ratios \cite{lid}. SN~1998bw rivals the brightest radio supernovae yet observed and radio observations indicate that the shock of the explosion was relativistic \cite{ku,wi}. Assuming the association and the low redshift \cite{ti} the GRB was also unusual -- at least 4 orders of magnitude fainter than other GRBs \cite{ga}. 

To what extent SNe are associated with GRBs is unclear due to the low numbers of well-observed Ib/c SNe and the large positional error of most GRBs. Wang \& Wheeler \shortcite{wa} argue that all Ib/c could produce GRBs, but due to beaming we would see only a fraction. Kippen et al. \shortcite{ki} found no evidence
for an association between SNe and strong GRBs, and Bloom et al. \shortcite{bl}
model the radio signature and suggest that 1\% of GRBs are produced by SNe. If  SN~1998bw is a member of a  previously unobserved subclass of GRB progenitors \cite{iw}, we have a rare opportunity to test and refine our understanding of SNe and GRBs. 

\subsection{Type Ib/c supernovae}

Supernovae of type Ib/c are identified by their early optical spectra, which lack the deep Si~II  absorption feature seen at 6150 \AA\ in Ia spectra and the prominent hydrogen lines of Type II SNe. Unlike Ia SNe, Ib/c objects are radio emitters and are typically fainter at maximum  by $M_{B} \sim 1.5$ magnitudes \cite{fi}. The parent galaxies of these events (Sbc or later) and heterogeneity of this class 
suggest that Ib/c SNe are powered by the same mechanism as Type II SNe -- core collapse of a massive star. The progenitors of Ib/c SNe have been modelled as Wolf-Rayet stars which have lost their hydrogen envelopes either via close binary interaction (Nomoto, Iwamoto \& Suzuki 1995), through a strong stellar wind (Woosley, Langer \& Weaver 1993) or a combination of the two mechanisms (Woosley, Langer \& Weaver 1995). 

This class is further divided into Ib SNe with strong helium lines in the early spectra, and Ic where helium lines are weak or absent. Opinion varies as to the relationship between Ic and helium-rich Ib SNe and whether there is a smooth or bimodal variation of He~I strengths \cite{fi2,cl2}. Ic progenitors may have lost their helium envelopes in another stage of mass loss, leaving a bare CO star at core collapse \cite{ha,no} or the helium envelope may be poorly mixed with the Ni$^{56}$ \cite{we}. Piemonte \shortcite{pi}
stresses the spectral and photometric variation for both Ib and Ic SNe.
These variations indicate a wide range of ejecta mass, which would depend on the main sequence mass of the progenitor and its mass loss history as well as secondary parameters such as metallicity and convection \cite{wlw2}. In general, ejecta mass is expected to be small compared to other SN types.

\subsection{SN~1998bw}

The models for SN~1998bw presented to date tend to fall into two classes -- an intrinsically energetic event or hypernova \cite{iw,wo} with a massive progenitor star and more normal SNe artificially brightened by beaming \cite{wa}. Radio observations suggest that material has been ejected irregularly  by a central engine \cite{li}. All models agree in requiring some form of non-symmetric geometry, possibly produced by an asymmetric explosion. 

With such diverse interpretations, it is important that all available data are used to provide observational limits for the models.
In this paper we present the results of a cooperative spectral monitoring campaign carried out at Siding Spring, Australia, on the AAT, UKST and SSO 2.3-m telescope between May and November 1998 and compare the spectral evolution and velocity shifts of SN~1998bw with other well-observed Ic SNe. 

\begin{figure}
\vspace*{-88mm}
\hspace*{32mm}
\vbox{
\centerline{
\psfig{figure=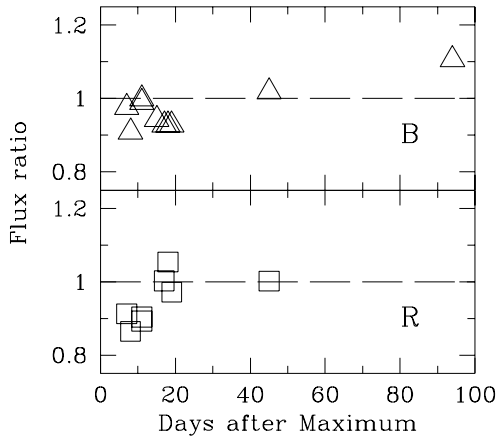,width=160mm}
}}
\vspace*{-15mm}
\caption{Comparison of $B$-band (top panel) and $R$-band (bottom panel) spectrophotometry from the corrected spectra with published photometry. Both bands show scatter with $\sigma$ = 6\%.}
\end{figure}

\section{Observations}

Spectral observations of SN~1998bw were made in Director's override time and service time on the Anglo-Australian Telescope (AAT) and UK Schmidt Telescope (UKST), and with the cooperation of scheduled observers on the Siding Spring Observatories (SSO) 2.3-m telescope. Unfortunately, the site experienced the poorest observing statistics of the decade and $>60$\% of allocated time was lost. We obtained useful spectra at 10 epochs which span 7 to 94 days past $V_{max}$ (or 23 to 110 days past the GRB event). Observational details are given in Table~1.

Observations were made using the scheduled instruments for the telescopes 
which included both conventional long-slit spectrographs (DBS, RGO) and fibre-fed spectrographs (2dF, FLAIR). The Nasmyth~B Imager at the 2.3-m was used as a long-slit spectrograph by inserting blue and red grisms and an order-sorting filter. Long-slit data were processed in the usual way using the FIGARO data reduction package \cite{sh}. A first order correction has been made for background emission from the galaxy, but narrow emission from underlying H~{\sc ii} regions remain in the spectra. At least two wavelength regions were observed on each of the long-slit data epochs, and spectra were combined by normalising to match the overlap regions. 

Data taken on FLAIR were processed using the IRAF data reduction package as outlined in Drinkwater \& Holman \shortcite{dr}. 2dF observations were extracted using the S-DIST and C-DIST utilities from FIGARO. Sky emission was removed using neighbouring fibres. It was not possible to correct the 2dF or FLAIR data for background galaxy emission, but the level of contamination was low during this period with the exception of the narrow nebular lines.

Telluric absorption lines have been removed from the data using observations of low-metallicity stars observed on days 7, 11 and 19, scaled where necessary to match the strongest O$_{2}$ features. Considering the poor conditions it is probably not surprising that telluric line ratios were variable and weak residuals can be seen in the spectra, particularly around 7500 to 8100~\AA\ and at $\lambda > 9500$~\AA.

\begin{figure}
\hspace*{9mm}
\vbox{
\centerline{
\psfig{figure=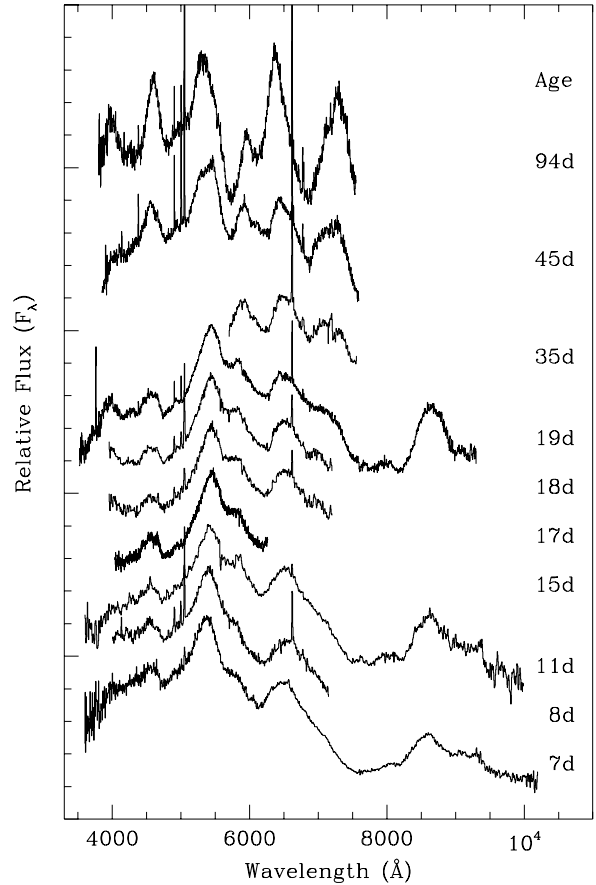,width=120mm}
}}
\caption{Spectral evolution of SN~1998bw. Ages are shown relative to date of visual maximum, 1998 May 12. Spectra have been normalised by the V-band photometry (see text) and offset in flux for the purpose of comparison.}
\end{figure}

Long-slit observations were corrected for instrumental response using a 
contemporaneous observation of a spectrophotometric standard. Fibre observations did not include accompanying standards, so a linear interpolation between bracketing long slit observations were used to correct these data. As these observations were made in non-photometric conditions and through narrow fibres or slits ($\sim$2 arcsec), correction to absolute flux was made by scaling to match the $V$-band light curve \cite{ga,mc} (the $V$-band was not covered on day 35 so the $R$-band was used). The success of this method was checked by comparing the photometry in the $B$ and $R$ bands with spectrophotometry from the corrected spectra, measured using filter responses from Bessell (private communication). Both $B$ and $R$ bands resulted in a mean ratio of 0.95, with $\sigma$ = 6\% (Figure 1). Errors at the edge of the spectra are likely to be larger and in general line fluxes from this data set should be regarded with caution. 
For the purpose of comparison, spectra shown in this paper have been normalised by the $V$-band photometry, and shifted to rest wavelengths using $vz = 2580$ km~s$^{-1}$ as derived from the narrow H~{\sc ii} region emission. Both the fluxed and normalised spectra are available via anonymous ftp at ftp.aao.gov.au/pub/local/ras/98bw.

\begin{figure*}
\begin{minipage}{175mm}
\vspace*{-47mm}
\vbox{
\centerline{
\psfig{figure=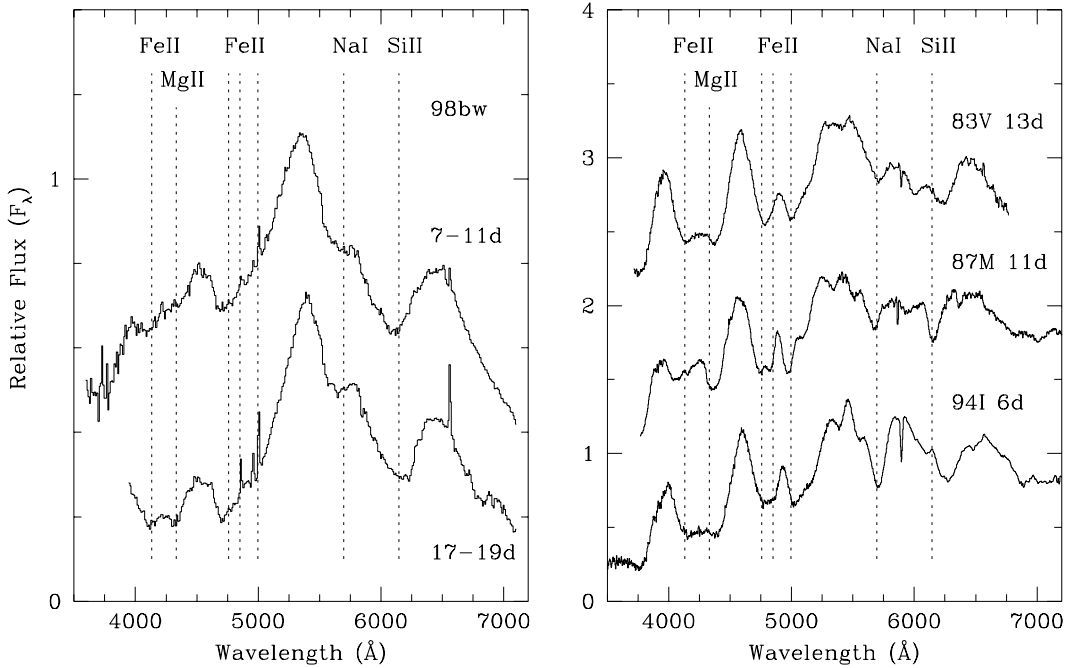,width=150mm}
}}
\vspace*{-13mm}
\caption{Left: SN~1998bw on days 7 to 11 and days 17 to 19. Right: Spectra of other Ic SNe at similar epochs. All spectra have been corrected for the redshift of the parent galaxy. Positions of median line centres shifted by $-10000$ km s$^{-1}$ relative to rest are shown in both panels for the Fe~{\sc ii} multiplet $\lambda$4274, Mg~{\sc ii} $\lambda$4481, Fe~{\sc ii} $\lambda \lambda$4923, 5018, 5169, Na~{\sc i}  $\lambda \lambda$5890, 5896 and Si~{\sc ii} $\lambda \lambda$6347, 6371. }
\end{minipage}
\end{figure*}

\begin{figure*}
\begin{minipage}{175mm}
\vspace*{-25mm}
\vbox{
\centerline{
\psfig{figure=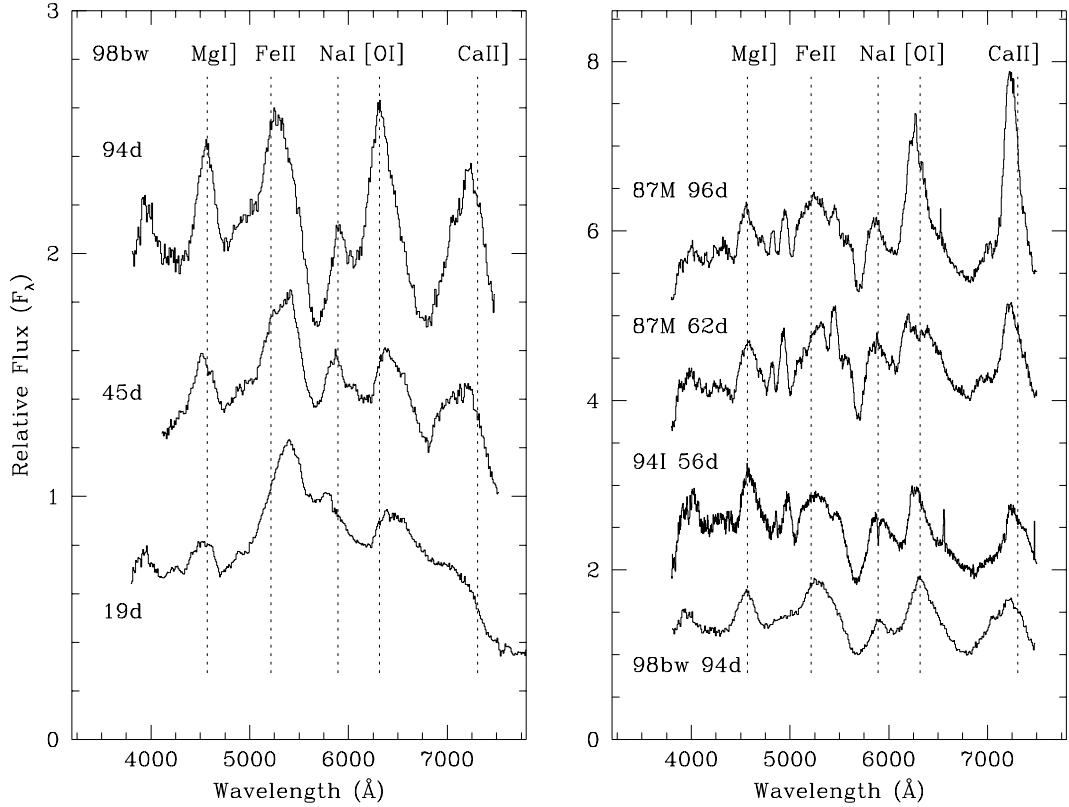,width=150mm}
}}
\vspace*{-13mm}
\caption{Left: Comparison of spectra taken on days 19, 45 and 94. Spectra have been normalised as in previous plots and shifted to rest at $z = 0.0086$. 
In addition, nebular lines from nearby H~{\sc ii} regions have been removed. Right: Spectra of other Ic SNe at similar epochs.
Median rest wavelengths are indicated for Mg~{\sc i}] $\lambda$4571, Fe~{\sc ii} $\lambda$5215, Na~{\sc i} $\lambda \lambda$5890, 5896, [O~{\sc i}] $\lambda \lambda$6300, 6363 and Ca~{\sc ii}] $\lambda \lambda$7298, 7323.}
\end{minipage}
\end{figure*}

\begin{table*}
\begin{minipage}{100mm}
\caption{Absorption minima measurements.}
\begin{tabular}{lccccc}
\hline
 Line & $\lambda_{r}$ & 98bw & 83V & 87M & 94I   \\
      & (\AA) & 19d  & 13d & 11d & 6d    \\
 & & & & & \\
Ca II & 3950 & $3750 \pm 20^{1}$ & $3760 \pm 20$ &   & $3760 \pm 10$     \\
      &      & $-15.2 \pm 1.5^{2}$ & $-14.4 \pm 1.6  $ &    & $-14.4 \pm 0.8$   \\ 
 & & & & & \\
Fe II & 4274 & $4130 \pm 30$   & $4150 \pm 20$   & $4060 \pm 30$   & $4170 \pm 30$   \\
      &      & $-10 \pm 2$ & $-8.7 \pm 1.6 $ & $-15 \pm 2$ & $-7 \pm 2 $  \\ 
 & & & & & \\
Mg II & 4481 & $4330 \pm 30$ & $4370 \pm 20$  & $4360 \pm 20$ & $4380 \pm 20$  \\
     &      & $-10 \pm 2$ & $-7.4 \pm 1.4$ & $-8.1 \pm 1.4$ & $-6.8 \pm 1.4$  \\
 & & & & & \\
Fe II  & 4555 & $4330 \pm 30$ & $4370 \pm 20$ & $4360 \pm 20$ & $4380 \pm 20$  \\
    &       & $-15 \pm 2$ & $-12.2 \pm 1.4$ & $-12.8 \pm 1.4$ & $-11.5 \pm 1.4$ \\
 & & & & & \\
Fe II  & 4923 & $4700 \pm 20$ & $4780 \pm 10$ & $4750 \pm 10$  & $4800 \pm 50$    \\
       &      & $-13.6 \pm 1.3$ & $-8.7 \pm 0.6$   & $-10.5 \pm 0.6$ & $-8 \pm 3$    \\
 & & & & & \\
Fe II  & 5018 & $4820 \pm 30$ & $4780 \pm 10$ & $4820 \pm 10$  & $4800 \pm 50$       \\
       &      & $-12 \pm 2$   & $-14.2 \pm 0.6$  & $-11.8 \pm 0.6$ & $-13 \pm 3$   \\
 & & & & & \\
Fe II  & 5169 & $4960 \pm 30$  & $5000 \pm 20$ & $4980 \pm 10$ & $5020 \pm 10$  \\
       &      & $-12 \pm 2$ &  $-9.8 \pm 1.2$  &  $-11.0 \pm 0.6$ & $-8.6 \pm 0.6$  \\
 & & & & & \\
Na I   & 5893 & $5660 \pm 30$ & $5710 \pm 10$ & $5677 \pm 5$ & $5710 \pm 7$       \\
       &      & $-11.9 \pm 1.5$ & $-9.3 \pm 0.5$ & $-11.0 \pm 0.3$ & $-9.3 \pm 0.4$ \\
 & & & & & \\
Si II  & 6357 & $6190 \pm 30$  & $6240 \pm 20$  & $6152 \pm 5$ & $6280 \pm 20$      \\
       &      & $-7.9 \pm 1.4$ & $-5.6 \pm 1$  & $-9.7 \pm 0.3$ & $-3.7 \pm 1.0$     \\
 & & & & & \\
O I    & 7774 &               &                & $7490 \pm 10$ &  $7560 \pm 20$    \\
       &      &               &                & $-11.0 \pm 0.4$ & $-11.4 \pm 1.1$   \\
\hline
\end{tabular}
$^{1}$Position of absorption minima in wavelength (\AA).\\
$^{2}$Relative velocity ($10^{3}$ km s$^{-1}$).
\end{minipage}
\end{table*}

\section{Spectral Evolution}

Early spectral evolution of SN~1998bw has been presented in Iwamoto et al. 
\shortcite{iw} (days --9 to +11). In Figure 2 the spectral evolution of 
SN~1998bw is shown between days 7 and 94. During this period, spectra are dominated by a strong continuum peaking around 5400 \AA, with a small number of  broad features which become increasingly dominant  relative to the continuum. Line widths remain approximately stable. Our observation on day 94 agrees qualitatively with the description by Patat \& Piemonte \shortcite{pa2} of the spectrum on day 123.

The breadth of the spectral features and the uncertainty in continuum level
hampers detailed analysis. In this paper we present preliminary results  by looking at the most notable features of the spectrum in comparison to classic Ic SNe SN~1983V \cite{cl}, SN~1987M \cite{fi3} and SN 1994I \cite{cl2,fi2}, and to SN 1997ef which has unusually broad lines and bears the closest resemblance to SN~1998bw \cite{iw2}.

\subsection{Days 7-19}

Eight of our spectra fall within the photospheric period, during which the B-band and V-band light curves were declining steeply prior to the radioactive tail. Spectra show little change over the period in the range 4000 to 7000~\AA, as seen in Figure~3 (left panel), where spectra have been binned in wavelength and time to improve the signal-to-noise ratio. The height of the continuum is poorly defined as overlapping P-Cygni profiles result in line blanketing over much of the spectral range -- apparent peaks are merely regions of relatively low opacity \cite{iw}. In the right panel of Figure~3 spectra of other supernovae are shown at similar epochs. Qualitively SN~1998bw is very different from the other supernovae, with merged absorption blueward and redward of 5300~\AA. SN~1998bw lacks the spectral detail and the overall effect is that
of heavy smoothing. However, detailed comparison between the SNe spectra show that the same bands can be identified, and most differences in SN~1998bw can be attributed to the large line widths.

In Table~2, the wavelengths of the observed minima in SN~1998bw (day 19) are compared with those measured from other SNe, and converted to relative velocities (in units of $10^{3}$ km s$^{-1}$) for suggested line identifications (see below). Estimates of measurement errors  are shown, reflecting how distinct the minima are for each line. SN~1998bw blueshifts are $\sim$30\% higher than SN~1983V and $\sim$50\% higher than SN~1994I, but are comparable with SN~1987M.  

The absorption band between 4100 -- 4300 \AA\ is resolved into two 
minima in SN~1998bw by days 17-19. The two minima are also seen in SN~1983V and
SN~1994I. This absorption band is present but not resolved in SN~1997ef. In SN~1987M the band extends blueward compared to the other SNe. The absorption band is attributed to Fe~{\sc ii} blends, centred at 
rest wavelengths of 4274~\AA\ and 4555~\AA\ \cite{cl}. An alternative identification for the second minimum is Mg~{\sc ii} $\lambda$4481 \cite{fi}, which gives velocities which are more in agreement with other features for all but SN~1987M. Other species identified in this region are Ti~{\sc ii} and C~{\sc ii} \cite{ba} and Cr~{\sc ii} \cite{iw}.

Fe~{\sc ii} is also the usual identification for the absorption band between 4700~\AA\
and 5200~\AA. In SN~1987M the feature is resolved into three minima which match well with Fe~{\sc ii} $\lambda \lambda$4923, 5018, 5169 (multiplet 42). In SN~1983V and SN~1994I
only two minima are resolved, identified as 5018~\AA\ and 5169~\AA\ \cite{cl,iw,iw2}. However, we find that the identification of the bluer line as 4923~\AA\ results in better agreement with other features. In SN~1998bw the band is only marginally resolved on day 19, with 4923~\AA\ unusually dominant. In SN~1997ef the line ratios are more typical, and the band is resolved into two minima \cite{iw2}.

In Ic SNe spectra, Na~{\sc i} $\lambda \lambda$5890, 5896 absorption typically strengthens later than the Fe~{\sc ii} features, becoming dominant around day 30 and fading by day 100. The line emerges more slowly in SN~1998bw compared to SN~1983V, SN~1987M and SN~1994I, but is comparable to SN~1997ef. The Na~{\sc i} blueshift is similar to 
Fe~{\sc ii} blueshifts for all four SNe in Table~2. The presence of He~{\sc i} $\lambda$5876 discussed for other Ic SNe \cite{cl2,cl} cannot easily be ascertained for SN~1998bw since the weaker line would be severely blended with Na~{\sc i}.

The Si~{\sc ii} $\lambda \lambda$6347, 6371 absorption shows the largest shift in velocity during this period, with minima velocities of $-11600$ km s$^{-1}$ on day 11 and 
$-7900$ km s$^{-1}$ on day 19. In Table~2, the Si~{\sc ii} line has the lowest blueshift of the measured lines for all SNe. The line profile of the Si~{\sc ii} feature is similar to that of SN~1983V, suggesting that there may be contribution from a second line resolved in SN~1983V, SN~1994I and SN~1997ef and
identified by Clocchiatti et al. \shortcite{cl} as O~{\sc i} $\lambda$6158.

\begin{figure}
\vspace*{-25mm}
\hspace*{13mm}
\vbox{
\centerline{
\psfig{figure=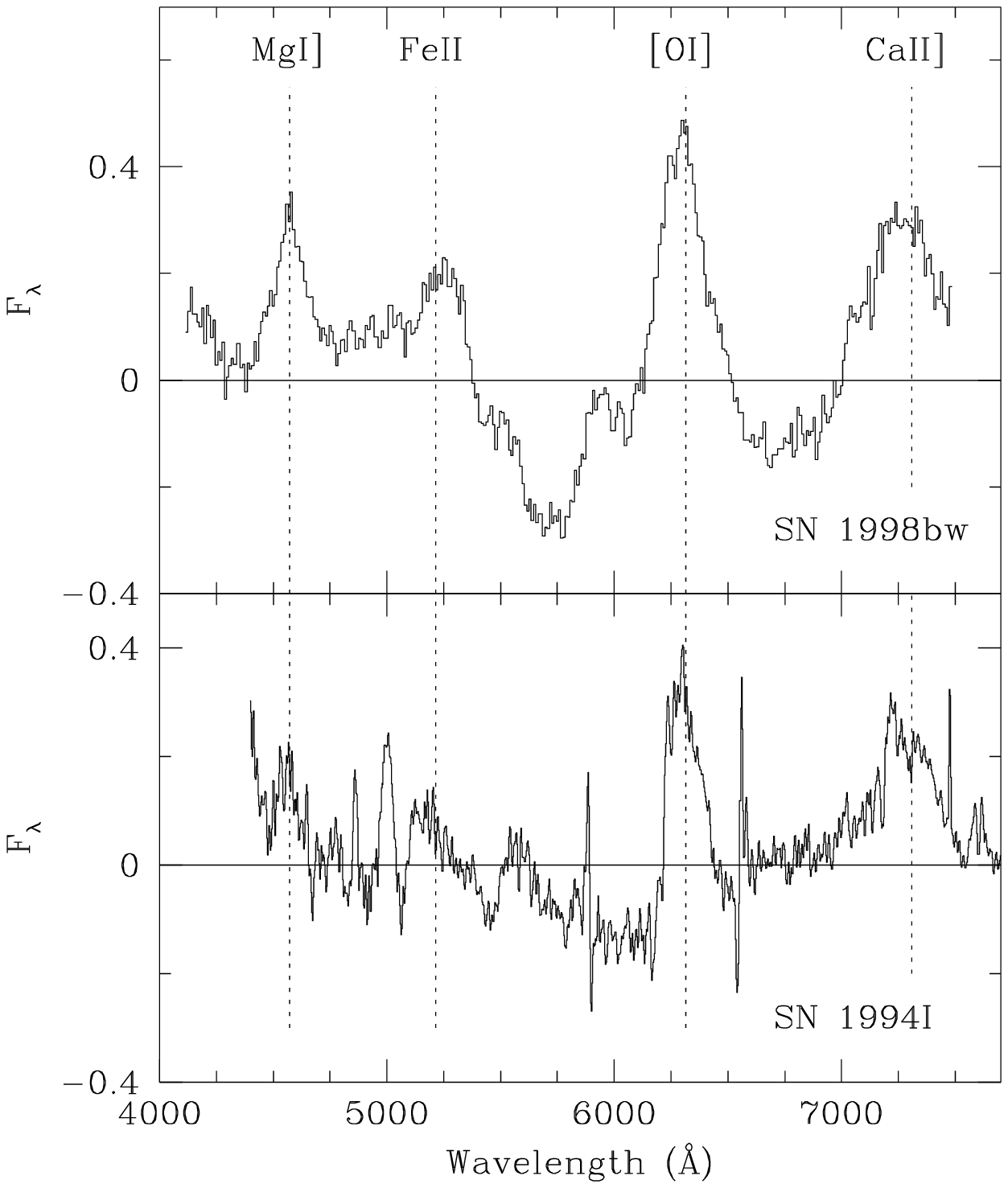,width=120mm}
}}
\vspace*{-5mm}
\caption{Top: Difference between spectra of SN~1998bw observed on day 45 and 94, after scaling by the V-band photometry. Narrow lines from nearby H~{\sc ii} regions have been removed. Excess emission from broad nebular lines are visible as SN~1998bw enters the supernebular phase. Bottom: Difference between day 36 and 56 spectra of SN~1994I.}
\end{figure}

\begin{table}
\caption{Velocity shifts and widths during the nebular phase.}
\begin{tabular}{lccccc}
\hline
Line & $\lambda_{r}$ & 98bw &  94I  &  87M  &  87M \\
     &               & (94d)  & (56d) & (62d) & (96d) \\
     &   &   &   &  &  \\
Mg~{\sc i}] & 4571 & $-1.4^{1}$ & $+0.9$ & $+0.2$ & $-1.0$ \\
            &      & $11.9^{2}$ &  8.4   & 8.6   & 7.3   \\
{[}O{\sc i}] & 6315  & $+0.7$ & $-1.2$ &  & $-1.9$ \\
            &        & 11.4   & 7.8    &    & 10.0 \\
Ca~{\sc ii}] & 7308 & $-4.0$ & $-1.6$ & $-2.7$ & $-2.6$ \\
             &      & 12.1   & 8.2   &  7.5    & 6.5    \\
Na~{\sc i}] & 5893$^{3 }$ & $-10.1$ & $-11.3$ & $-10.0$ & $-9.9$ \\
            &            & 11.1    & 9.0     &  5.5    & 4.5  \\
    &   &   &     &    &   \\ 
\hline 
\end{tabular}
 \\
$^{1}$Velocity shift of line centre ($10^{3}$ km s$^{-1}$). \\
$^{2}$Width of features at half height ($10^{3}$ km s$^{-1}$). \\
$^{3}$Absorption component. \\
\end{table}

\subsection{Days 19-94}

Late-time spectral evolution of SN~1998bw is shown in Figure~4 (left panel). During this period the light curve is decaying linearly via radioactive decay \cite{mc}, and as expected we see a transition from an absorption-dominated photospheric spectrum to an emission-dominated nebular spectrum, though on day 94 the transition is still not complete. Early nebular spectra of SN~1994I and SN~1987M are shown in comparison to SN~1998bw in Figure~4 (right panel). SN~1998bw evolves slowly compared to the other two SNe  -- the day 94 spectrum bears a closer resemblance to SN~1987M on day 62 than day 96, and is very similar to SN~1994I on day 56. In order to separate the new emission from the other features in SN~1998bw, the day 45 spectrum has been subtracted from the day 94 spectrum, after scaling by the V-band photometry. Likewise, day 36 has been subtracted from day 56 for SN~1994I. The results are compared in Figure~5. 

Velocity shifts of the main features are given in Table 3.
[O{\sc i}] $\lambda \lambda$6300, 6363 is blueshifted in SN~1987M and SN~1994I. In SN~1998bw on day 94 the line profile is 
symmetric, but in the residual emission it is also blueshifted by $\sim -500$ km~s$^{-1}$. Ca~{\sc ii}] $\lambda \lambda$7291, 7323 has a similar blueshift in SN~1987M and SN~1994I. In SN~1998bw the blueshift is significantly greater, but
in the residual spectrum the profile is similar to SN~1994I, so apparent differences are probably due to contribution from the persisting photospheric features. The Ca~{\sc ii}] to [O{\sc i}] ratio is greater in SN~1987M than in 
SN~1994I and SN~1998bw, which has been attributed to a difference in the relative abundances of calcium and oxygen \cite{fi2}. 
While Mg~{\sc i}] $\lambda$4571 is the usual identification for the emission peak at 4500~\AA\ \cite{fi}, an alternative identification is given by Patat et al. \shortcite{pa2} as Fe~{\sc ii} $\lambda$4555, and both transitions give an adequate fit to the line with a symmetric profile. We adopt the Mg~{\sc i}] identification for this paper. 

Approximate line widths have been measured using ABLINE in FIGARO (Table~3). Line widths are similar for the Mg~{\sc i}], [O~{\sc i}] and Ca~{\sc ii}] emission for each supernova.  The  mean width of these lines in SN~1998bw on day 94 is $11600 \pm 400$ km~s$^{-1}$ which is $\sim$45\% broader than SN~1987M and SN~1994I. Na~{\sc i} emerges as a P-Cygni profile between days 19 and 94, evolving more slowly in equivalent width and in emission to absorption ratio than SN~1987M and SN~1994I. A noticable difference between SN~1998bw and SN~1987M is the width of the absorption component of Na~{\sc i}, which is only half the width in SN~1987M, and considerably narrower than the emission lines. 

The peak at 5200 \AA\ has been tentatively identified as Fe~{\sc ii} $\lambda$5215 \cite{pa2}. This transition is typically seen in older Ia SNe, but not in Ic SNe, and Patat and Piemonte note that its presence would indicate that SN~1998bw was a type Iac SN. However, the feature is also present in SN~1987M
and more weakly in SN~1994I during the early nebular phase. It fades 
relative to Mg~{\sc i}], [O~{\sc i}] and Ca~{\sc ii}] at later times. It therefore seems reasonable to assume that the presence of Fe~{\sc ii} emission in Ic SNe spectra is typical during the transition from the photospheric to the nebular phase.
In the residual spectrum (Figure~5) the feature is weaker relative to the 4500~\AA\ peak and has a markedly different profile. This supports the
identification of the 4500 and 5200~\AA\ peaks as transitions of different species, and indicates that the 5200~\AA\ peak is fading and/or is artificially
enhanced by a relatively high continuum level.

\begin{figure}
\vspace*{-15mm}
\hspace*{12mm}
\vbox{
\centerline{
\psfig{figure=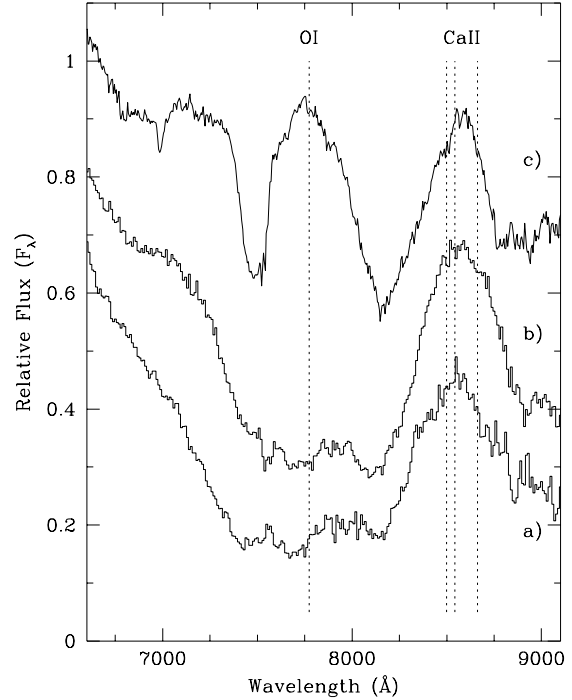,width=120mm}
}}
\vspace*{-12mm}
\caption{The 7000 to 9000 \AA\ region of Ic SNe spectra is dominated by 
the O~{\sc i} $\lambda$7774 multiplet and Ca~{\sc ii} $\lambda \lambda$8498, 8542, 8662 (rest wavelengths are shown). These lines are blended in SN~1998bw on day 11 (a) and day 19 (b), but are well separated in SN~1987M on day 7 (c). Note that weak residual telluric lines are present in the SN~1998bw spectra, especially around 7600~\AA.}
\end{figure}

\subsection{7000 -- 9000 \AA}

The red spectral region of SN~1998bw is shown in Figure~6 for days 7 and 19. It is this region which differs most markedly from typical Ic SNe such as SN~1987M (shown on day 7). In these SNe, strong absorption 
from O~{\sc i} $\lambda$7774 and Ca~{\sc ii} $\lambda \lambda$8498, 8542, 8662 are well separated by an absorption free region missing in SN~1998bw. 
Iwamoto et al \shortcite{iw} have successfully modelled this region for day~$-9$
with a photospheric velocity of 28000 km~s$^{-1}$. However, by day~$-1$ their model predicts that the two features should be resolved, and that by day~7 O~{\sc i} and Ca~{\sc ii} are unblended. Similar evolution is predicted by the direct analysis models of Branch \shortcite{br}. In SN~1997ef this behaviour is seen, with blended absorption on day~3 and well separated features on day~30 \cite{iw2}. In SN~1998bw, however, there is no sign of significant separation as late as day~19, our last epoch covering this region.  

While it is presumably possible to fit this region by increasing the mass of the progenitor, and therefore the line widths
of the O~{\sc i} and Ca~{\sc ii} profiles, there are two limitations which need to be considered. The first is that we do not expect to see redshifted absorption from a simple expanding envelope. Under this assumption, 
O~{\sc i} cannot contribute to the absorption band redward of 7774~\AA, irrespective of the line width. The second limitation is that we detect a shallow dip at $\sim$8100~\AA\ which we identify as the minimum of the Ca~{\sc ii} absorption. This feature aligns well with the Ca~{\sc ii} $\lambda \lambda$3933, 3968 line profile, and the velocity of $-15300$ km~s$^{-1}$ is already higher than blueshifts of other lines at this epoch (Table~3). 

In order to reproduce the spectrum on day 19 we would require a highly
unusual geometry to produce redshifted absorption from O~{\sc i}, or to produce
Ca~{\sc ii} absorption with a second minimum at around $-24000$ km~s$^{-1}$.
An alternative explanation is the presence of a third component absorbing at around 7700 -- 8000 \AA. Adequate fits can be produced with lines of rest wavelength 8000 -- 8350 \AA, and candidate species include C~{\sc ii}, C~{\sc iii} and N~{\sc i}. A relatively weak contribution from any of these species would result in the blended spectrum we observe. Modelling is required to further investigate this region, and to determine whether unusual abundances, density or temperature distribution can explain the observations. 

\section{Conclusion}

During the period  between 7 and 94 days after V-band maximum, we have seen that SN~1998bw resembles other Ic SNe sufficiently to support this classification, but has unusually slow spectral evolution. On day 94 we see the emergence of a nebular spectrum, which retains many of the characteristics of the photospheric period. The late onset of the supernebular phase, compared to SN~1987M (62~d)
and SN~1994I (56~d), is consistent with the ejection of an unusually large mass, as predicted by lightcurve models \cite{iw,wo}. 

By day 19, SN~1998bw blueshifts are up to 50\% larger than other Ic SNe. However, increased blueshifts alone seem insufficient to explain the unusually smooth and blended spectrum which persists to late times -- for instance SN~1998bw on day 19 has similar blueshifts to SN~1987M on day 11, but SN~1987M has well defined spectral features.
Emission line widths on day 94 are 45\% broader than SN~1994I and SN~1987M and Na~{\sc i} absorption is far broader in SN~1998bw on day 94 than in SN~1987M at similar epochs. The line profiles of SN~1998bw may have disproportionally strong absorption wings -- we lack an example of an unblended feature at earlier times for confirmation. Using the standard model for a homologously expanding envelope, absorption lines which are broader but of similar blueshift imply that the absorption region in SN~1998bw spans a larger range of velocity space, at both higher and lower velocities than other Ic SNe, as expected for a massive envelope. A closer inspection of the 7000 -- 9000 \AA\ region of SN~1998bw suggests that we are also seeing contribution from enhanced line species. Unusually strong lines from species such as N~{\sc i}, C~{\sc ii}, C~{\sc iii}, Ti~{\sc ii} and Cr~{\sc ii} may help to produce the extensive line blending.
If so, this could indicate an overabundance of these elements, or unusual physical properties of the ejecta.

More work is required in establishing line identifications and spectral characteristics, best done using spectral models. Whether or not SN~1998bw was
associated with GRB 980425, it is of great interest as an extreme example of Ic SNe. SN~1997ef bears some resemblance to SN~1998bw and may be an intermediate object between SN~1998bw and classical SNe, though it is too early to say whether we are seeing a bimodal or continuous variation in properties. Thanks to the interest inspired by the $\gamma$-ray burst, SN~1998bw has been observed extensively and successful modelling of this object is likely to enhance our understanding of this relatively poorly observed class of supernova.

\section*{Acknowledgements}

We thank the observers at the AAT, UKST and MSO 2.3-m telescope who kindly donated their time to our project. We thank Alejandro Clocchiatti, Craig Wheeler, Jodie Martin and Paolo Mazzali for providing digital supernova data to compare with our observations, and Peter Meikle, David Branch and Brian Schmidt for helpful discussions.


\begin{thebibliography}{}
\bibitem[\protect\citename{Baron et al.\ }1996]{ba}
  Baron E., Hauschildt P. H., Branch D., Kirshner R. P., Filippenko A. V.,
  1996, MNRAS, 279, 799, astro-ph/9510070
\bibitem[\protect\citename{Bloom et al.\ }1998]{bl}
  Bloom J. S., Kulkarni S. R., Harrison F., Prince T., Phinney E. S., 1998, 
  ApJ, 506, L105, astro-ph/9807050
\bibitem[\protect\citename{Branch }2000]{br}
  Branch D., 2000, in Livio M., ed, Supernovae and Gamma-Ray Bursts, in press, astro-ph/9906168
\bibitem[\protect\citename{Clocchiatti et al.\ }1996]{cl2}
  Clocchiatti A., Wheeler J. C., Brotherton M. S., Cochran A. L., Wills D.,
  Barker E. S., Turatto M., 1996, ApJ, 462, 462
\bibitem[\protect\citename{Clocchiatti et al.\ }1997]{cl}
  Clocchiatti A. et al., 1997, ApJ, 483, 675
\bibitem[\protect\citename{Drinkwater \& Holman }1996]{dr}
  Drinkwater M., Holman B., 1996, FLAIR Data Reduction with IRAF, 
  http://www.aao.gov.au/local/www/cgt/flair\_iraf /flair\_iraf.html
\bibitem[\protect\citename{Filippenko }1997]{fi}
  Filippenko A. V., 1997, ARA\&A, 35, 309
\bibitem[\protect\citename{Filippenko et al.\ }1995]{fi2}
  Filippenko A. V. et al., 1995, ApJ, 450, L11
\bibitem[\protect\citename{Filippenko, Porter \& Sargent }1990]{fi3}
  Filippenko A. V., Porter A. C., Sargent W. L. W., 1990, AJ, 100, 1575
\bibitem[\protect\citename{Galama et al.\ }1998]{ga}
  Galama T. J. et al., 1998, Nature, 395, 670, astro-ph/9806175
\bibitem[\protect\citename{Harkness et al.\ }1987]{ha}
  Harkness R. P. et al., 1987, ApJ, 317, 355
\bibitem[\protect\citename{Iwamoto et al.\ }1998]{iw}
  Iwamoto K. et al., 1998, Nature, 395, 672, astro-ph/9806382
\bibitem[\protect\citename{Iwamoto et al.\ }2000]{iw2}
  Iwamoto K., Nakamura T., Nomoto K., Mazzali P. A., Danziger I. J., Garnavich
  P., Kirshner R., Jha S., Balam D., Thorstensen J., 2000, ApJ, in press, 
  astro-ph/9807060 
\bibitem[\protect\citename{Kippen et al.\ }1998]{ki}
  Kippen R. M. et al., 1998, ApJ, 506, L27, astro-ph/9806364
\bibitem[\protect\citename{Kulkarni et al.\ }1998]{ku}
  Kulkarni S. R. et al., 1998, Nature, 395, 663, astro-ph/9807001
\bibitem[\protect\citename{Li \& Chevalier }1999]{li}
  Li Z.-Y., Chevalier R. A., 1999, ApJ, 526,716, astro-ph/9903483
\bibitem[\protect\citename{Lidman et al.\ }1998]{lid}
  Lidman C., Doublier V., Gonzalez J. -F., Augusteijn T., Harnaut O. R., 
  Boehnhardt H, Patat F., Leibundgut B., 1998, IAUC 6895
\bibitem[\protect\citename{McKenzie \& Schaefer }1999]{mc}
  McKenzie E. H., Schaefer B. E., 1999, PASP, 111, 964, astro-ph/9904397
\bibitem[\protect\citename{Nomoto et al.\ }1995]{no}
  Nomoto K., Iwamoto K., Suzuki T., 1995, Physics Reports 256, 173
\bibitem[\protect\citename{Patat \& Piemonte }1998a]{pa}
  Patat F., Piemonte A., 1998a, IAUC 6918
\bibitem[\protect\citename{Patat \& Piemonte }1998b]{pa2}
  Patat F., Piemonte A., 1998b, IAUC 7017
\bibitem[\protect\citename{Piemonte }2000]{pi}
  Piemonte A., 2000, in Phillips M., ed, SN~1987A: Ten Years Later, in press 
\bibitem[\protect\citename{Shortridge et al.\ }1997]{sh}
  Shortridge K., Meyerdierks H., Currie M., Clayton M., 1997, PPARC Starlink
  User Note, 86.13
\bibitem[\protect\citename{Tinney et al.\ }1998]{ti}
  Tinney C., Stathakis R., Cannon R., Galama T., 1998, IAUC 6896
\bibitem[\protect\citename{Wang \& Wheeler }1998]{wa}
  Wang L., Wheeler J. C., 1998, ApJ, 504, L87, astro-ph/9806212
\bibitem[\protect\citename{Wieringa et al.\ }1998]{wi}
  Wieringa M., Frail D. A., Kulkarni S. R., Higdon J. L., Wark R., 
  Bloom J. S., BeppoSAX GRB Team, 1998, IAUC 6896
\bibitem[\protect\citename{Woosley \& Eastman }1997]{we}
  Woosley S. E., Eastman R. G., 1997, in Ruiz-Lapuente R., Canal R.,
  Isern I., eds, Thermonuclear Supernovae, NATA ASI series, 821 
\bibitem[\protect\citename{Woosley et al. }1999]{wo}
  Woosley S. E., Eastman R. G., Schmidt B. P., 1999, ApJ, 516, 788, astro-ph/9806299
\bibitem[\protect\citename{Woosley et al.\ }1993]{wlw1}
  Woosley S. E., Langer N., Weaver T. A., 1993, ApJ, 411, 823
\bibitem[\protect\citename{Woosley et al.\ }1995]{wlw2}
  Woosley S. E., Langer N., Weaver T. A., 1995, ApJ, 448, 315 
\end{thebibliography}
\end{document}